\documentclass[aps,prl,twocolumn,groupedaddress,showpacs]{revtex4}%
\usepackage{amsfonts}
\usepackage{graphicx}
\usepackage{amsmath}
\usepackage{amssymb}%
\begin{document}
\title{Coulomb blockade in graphene nanoribbons}
\author{F. Sols$^{1}$, F. Guinea$^{2}$, and A. H. Castro Neto$^{3}$}
\affiliation{$^{1}$ Departamento de F\'{\i}sica de Materiales, Universidad Complutense de
Madrid, E-28040 Madrid, Spain }
\affiliation{$^{2}$ Instituto de Ciencia de Materiales, CSIC, E-28049 Cantoblanco, Madrid,
Spain }
\affiliation{$^{3}$ Physics Department, Boston University, 590 Commowealth Avenue, Boston,
Massachussetts 02215, USA}


\begin{abstract}
We propose that recent transport experiments revealing the existence of an
energy gap in graphene nanoribbons may be understood in terms of Coulomb
blockade. Electron interactions play a decisive role at the quantum dots which
form due to the presence of necks arising from the roughness of the graphene
edge. With the average transmission as the only fitting parameter, our theory
shows good agreement with the experimental data.

\end{abstract}
\volumeyear{year}
\volumenumber{number}
\issuenumber{number}
\eid{identifier}
\startpage{1}
\endpage{ }
\maketitle



\bigskip

Graphene, a two-dimensional allotrope of carbon on a honeycomb lattice, was
isolated a few years ago \cite{novoselov_04} creating a great excitement in
the physics community due to its close connections to high-energy particle
physics \cite{pw_06,kats_nat06} and its tantalizing possible technological
applications \cite{fujita,berger_06}. It is now experimentally established
that a great deal of the properties of graphene \cite{geim_rev} can be
described in terms of non-interacting (or weakly interacting) linearly
dispersing Dirac quasi-particles \cite{vozmediano,peres_prb06}. The only
accepted exception maybe when graphene is subject to strong magnetic fields in
the quantum Hall regime \cite{zhang}, when the electronic kinetic energy is
quenched by the appearance of Landau levels, and the fourfold degeneracy of
the Landau levels is split by electron-electron interactions
\cite{nomura,goerbig}. Nevertheless, in the absence of an applied magnetic
field, because of the vanishing of the density of states for Dirac fermions in
two dimensions \cite{wallace}, the electrons in graphene interact through
strong, essentially unscreened, long-range Coulomb interactions
\cite{shung,sols,sarma,macdonald}. The fact that Coulomb interactions do not
show up in bulk experiments remains a puzzle in the physics of graphene.

Recent experiments on the electron transport properties of lithographically
patterned graphene nanoribbons have shown the existence of an energy gap near
the charge neutrality point \cite{HOZK07}. The size of the gap $E_{g}$ is
inferred from the nonlinear conductance at low temperatures and is found to
decrease with the ribbon width $W$ following the approximate law $E_{g}%
=\alpha/(W-W^{\ast})$. This result seems to correlate with a conductance
behavior $G=\beta(W-W_{0})$, since $W^{\ast}\approx W_{0}\simeq16$ nm for the
same sample at temperature $T=1.6$ K. In the absence of interaction effects,
the energy gaps between subbands in a graphene ribbon should scale inversely
with the ribbon width, $E_{g}\approx\hbar v_{\mathrm{F}}/W$, where
$v_{\mathrm{F}}\approx10^{6}\,\mathrm{m/s}\approx0.66$ eV $\times$ nm
$\times\,\hbar^{-1}$ is the Fermi velocity in graphene \cite{geim_rev}.
Nevertheless, this estimate leads to gaps which are smaller than those
observed experimentally (note that, for widths of about 20 nm the experimental
gaps are larger than 0.1-0.2 eV). This result has led to the suggestion that
the effective transport width is reduced with respect to the nominal width $W$
by an amount $W^{\ast}\approx W_{0}$ due to the existence of structural
disorder at the edges or to a systematic inaccuracy in the determination of
the geometrical width caused by over-etching beneath the etch mask
\cite{HOZK07}. On the other hand, it has been shown that graphene quantum dots
as large as 25 $\mu$m (at low temperatures), and as small as 40 nm (at room
temperature), show Coulomb blockade effects \cite{geim_rev} indicating that
electron interactions become stronger as the dimensions of graphene sheets are reduced.

\begin{figure}[ptb]
\includegraphics[width=0.7\columnwidth]{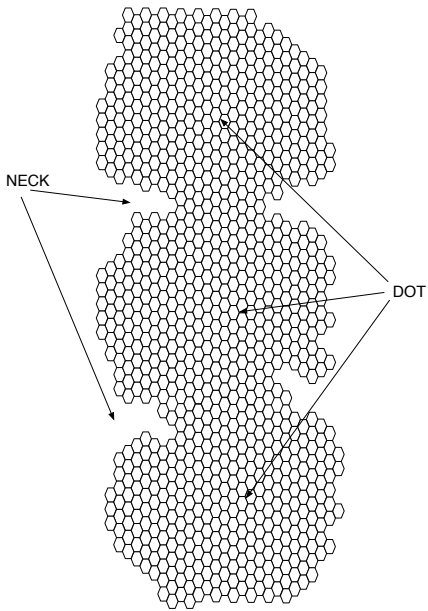}\caption{Illustration of a
graphene ribbon with a disordered edge leading to the formation of necks and
dots along the ribbon. Coulomb blockade takes place when the charge moves from
dot to dot. }%
\label{neckdot}%
\end{figure}

In the present work we argue that the main results of Ref. \cite{HOZK07} can
be naturally explained as due to Coulomb blockade effects originated by the
roughness at the edges of graphene nanoribbons. This roughness occurs
naturally in graphite samples, as has been seen in scanning tunneling
microscopy \cite{niimi}, leading to the localization of charge at the edges
and the formation of electronic puddles \cite{cn_hall}. As depicted in Fig.
\ref{neckdot}, disorder at the edges of a graphene ribbon also leads to the
formation of ``necks'', causing an abrupt reduction in the number of
conducting channels and thus to a large increase in the impedance along the
graphene sheet. This results in the electric isolation of nanoscale size
regions, or ``dots'', where the electrons become temporally confined. Within
this picture, Coulomb blockade \cite{BMS83,H83,BG85,AL91,GD92} results from
the electron transport from dot to dot through graphene necks.

In our analysis, we will follow the standard theory of Coulomb blockade
effects, neglecting the electronic level spacing within the grains. Note that
graphene states delocalized throughout a region of linear size $W$ show level
spacings of order $\hbar v_{\mathrm{F}} / W$, which is the same scaling
behavior followed by the charging energy of a grain of size $W$, $e^{2} / W$.
It seems likely, however, that the rough edges of the samples studied in
\cite{HOZK07}, as well as the internal lattice defects, can lead to a variety
of partly or fully localized states at energies close to the Fermi level,
reducing the electronic level spacing \cite{fujita,FWNK96,VLSG05,PGLPN06}.

In the presence of Coulomb blockade characterized by a renormalized charging
energy $E_{c}^{\ast}$ (which here we assimilate to $E_{g}$), the conductance
between neighboring metallic dots is renormalized to lower values as the
energy or temperature scale is reduced \cite{GS86,SZ90,HZS99}. This
renormalization begins to be appreciable at temperatures $T\lesssim
E_{c}/k_{B}$, where $E_{c}$ is the charging energy associated to the geometric
capacitance of the dot. A perturbative analysis, valid for relatively high
temperatures such that $T\sim E_{c}/k_{B}$, leads to a conductance given by:
\begin{equation}
G=G_{0}-G_{Q}c\ln\left(  \frac{E_{c}}{k_{B}T}\right)  ~,\label{G-ren}%
\end{equation}
where $c=2$ and $G_{0}=4G_{Q}N_{\text{ch}}\tau$ is the non-interacting
conductance (we include spin and valley degeneracy), $G_{Q}=e^{2}/h=38.7$
$\mu$S being the quantum of conductance, $N_{\text{ch}}=k_{F}W/\pi$ the number
of transverse orbital channels, and $\tau$ the average transmission per
channel. Eq. (\ref{G-ren}) is valid when the transmission per channel is low,
$\tau\ll1$, although the total conductance can take arbitrary values. A
similar expression can be obtained in the limit $\tau\lesssim1$, provided that
none of the channels has perfect transmission \cite{N99}. In the latter case,
the constant $c$ in Eq. (\ref{G-ren}) takes the value $c=8/\pi^{2}$. Within
the relatively narrow range of widths and temperatures considered in Ref.
\cite{HOZK07}, the logarithm in Eq. (\ref{G-ren}) can be taken as a constant
of order unity, leading to the approximate expression:
\begin{equation}
G\simeq(4G_{Q}k_{F}\tau/\pi)(W-W_{0})~,\label{G_W}%
\end{equation}
where $W_{0}\approx\pi c/4k_{F}\tau$.

The experimental dependence of $G$ on $W$ can be used to estimate the average
transmission $\tau$. Figure 2 of Ref. \cite{HOZK07} indicates that, at room
temperature, and for a gate voltage $V_{g}-V_{\mathrm{Dirac}}=-50$ V, a change
in conductance $\Delta G\approx80~\mu$S takes place if the width changes by
$\Delta W\approx40$ nm. This yields $W_{0} \approx40$ nm. The corresponding
hole density in this experiment is $n = 3.6 \times10^{12}$ cm$^{-2}$, which
implies that $k_{\mathrm{F}}^{-1} = 1/\pi\sqrt{n} \approx1.7$ nm. These
experimental results are consistent with Eq. (\ref{G_W}) if the average
transmission per channel is $\tau\approx0.07$. The total number of channels,
for a width $W \approx40$ nm, is $4N_{\mathrm{ch}} \approx4 k_{\mathrm{F}}
W/\pi\approx30$. The average transmission probability found here is consistent
with tight-binding calculations for wedge shaped graphene constrictions
\cite{MJFP06} (see also \cite{LVGC07,MB07}).

The analysis which at high temperatures leads to Eq. (\ref{G-ren}) also shows
that, at low temperatures, the effective charging energy is renormalized by
virtual jumps of the electrons across the junction, leading to:
\begin{align}
E_{g} &  \approx E_{c}e^{-G/cG_{Q}}\nonumber\\
&  =E_{c}e^{-4N_{\mathrm{ch}}\tau/c}~=E_{c}e^{-4k_{\mathrm{F}}W\tau/\pi
c}\,,\label{G_W_0}%
\end{align}
where $E_{c}$ is the charging energy for the completely isolated dot. It seems
reasonable to assume that $E_{c}$ is determined by the nominal ribbon width
$W$, $E_{c}\sim e^{2}/W$ since that is the size of a typical puddle which is
isolated from the rest of the ribbon through a contact (see Fig.
\ref{neckdot}). This contact acts as the bottleneck which determines the
conductance of the graphene nanoribbon. This fact does not preclude, however,
the possibility of further structure in the I-V characteristic which could be
induced by the presence of other dots.

The capacitance of the grains is also modified by the presence of metallic
leads and gates. The leads, and the regions of the quasi--one-dimensional
ribbon at distances greater than $W$ from the island considered do not change
appreciably the charging energy, as a one-dimensional charge distribution does
not screen an electrostatic potential. The screening of a metallic gate at a
distance $d$ from the island can be analyzed, when $d \gg W$, by assuming that
the charge in the island induces an image charge. The charging energy is
changed to $E_{c} \sim e^{2} / W - e^{2} / 2 d $. In the following, we neglect
the second term, as $W \sim20 - 100$ nm and typical distances to the gate
\cite{novoselov_04} are $d \sim300$ nm. Charging effects should be strongly
suppressed when the distance to the metallic gate is comparable, or smaller,
than the width of the ribbon.

Finally, we have identified the value of the Coulomb gap at low temperatures
with the gap in the I-V characteristics measured in Ref.~\cite{HOZK07}. Note
that the argument in the exponent in Eq. (\ref{G_W_0}) becomes $4k_{\mathrm{F}%
} W \tau/\pi c \approx0.8$ for $W \approx40$ nm, which leads to an appreciable
renormalization of the geometrical charging energy by virtual charge
fluctuations (see \cite{Getal00} for similar effects in a different granular
system). We note that $E_{c} \sim4$ meV for $W \approx40$ nm.

\begin{figure}[ptb]
\includegraphics[width=1\columnwidth]{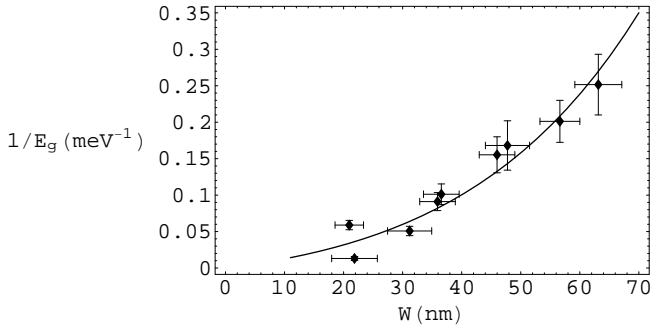}\caption{Comparison between
experimental data of Ref. {\cite{HOZK07}} and the theoretical result (full
line): $E_{g}^{-1} = B W e^{C W}$, obtained from Eq. ({\ref{EgW}}) with $B =
0.001 \, (\mathrm{meV \times nm})^{-1}$ and $C = 0.023 \, \mathrm{nm}^{-1}$. }%
\label{gap_fig}%
\end{figure}

We can also write Eq. (\ref{G_W_0}) as:%
\begin{equation}
E_{g}(W)\approx\frac{e^{2}}{W}e^{-W/W_{0}}~,~\label{EgW}%
\end{equation}
where $W_{0}$ is the length scale used in Eq. (\ref{G_W}). Thus, as a function
of $W$, $E_{g}^{-1}$ takes low values for $W\ll W_{0}$ while experiencing a
sharp rise for $W\agt W_{0}$. We note that, $E_{g}^{-1}(W_{0})\approx
5\times10^{-3}$ (meV)$^{-1}$, substantially smaller that the unit scale used
in Fig. 3e of Ref. \cite{HOZK07}. One can see from Fig.~\ref{gap_fig} that Eq.
(\ref{EgW}) explains the experimental data of Ref. \cite{HOZK07} within its
error bars. A comparison between the expression $E_{g}^{-1}=BWe^{CW}$,
obtained from Eq. (\ref{EgW}) and the data in \cite{HOZK07} yields
$C^{-1}=43\,\mathrm{nm}$, which is in good agreement with our theoretical
estimate for $W_{0}$. One also obtains $B=10^{-3}\,(\mathrm{meV\times
nm})^{-1}$, which agrees reasonably with $e^{-2}=6.95\times10^{-4}%
\,(\mathrm{meV\times nm})^{-1}$. The fact that $B^{-1}$ is smaller than
$e^{2}$ can be partly attributed to the screening effect of the gate.

Finally, we note some trends which provide additional qualitative support to
the Coulomb blockade picture. Figs. 3a, 3c, and 3d of Ref. \cite{HOZK07} show
the differential conductance as a function of the bias and gate voltage.
Electron-hole symmetry explains the symmetric behavior around a value of
$V_{g}$ which must be identified with the neutrality point. The maximum
vertical width of the dark (low differential conductance) zone must be
identified with the gap $E_{g}(W)$, which clearly decreases with $W$, in
qualitative agreement with Eq. (\ref{EgW}). On the other hand, varying $V_{g}$
is equivalent to changing $k_{\mathrm{F}}$. Eq. (\ref{G_W_0}) shows that the
variation of the gap and the differential conductance as a function of
$k_{\mathrm{F}}$ must be faster for large values of $W$, in good agreement
with the experimental results.

The experiment of Ref. \cite{HOZK07} also shows that the linear ($V_{b}
\rightarrow0$) conductance depends weakly on the gate voltage at the
neutrality point, where $k_{\mathrm{F}} \rightarrow0$. Our analysis predicts
that the linear conductance depends on gate voltage through the product
$k_{\mathrm{F}} \tau$. The well-known existence of a minimum in the bulk
conductivity can be translated, within a Drude picture, into $k_{\mathrm{F}}
\tau$ tending to a constant value as $k_{\mathrm{F}} \rightarrow0$. This trend
is consistent with the experimental observation described above.
The insensitivity of the conductance to the gate voltage can also be expected
to occur when puddles are formed with a finite (positive or negative) charge
density \cite{NK07,Metal07,GKV07}.

The analysis presented so far describes the observed low temperature gap in
transport measurements in terms of the features of isolated junctions,
neglecting effects associated to interference effects between multiple
junctions. We do not expect these effects to change significantly the
analysis. Coulomb blockade leads to the suppression of phase coherence between
successive tunneling events. The scaling equation (\ref{G-ren}) is valid for
granular arrays, and transport gaps in these systems have the same functional
dependence as in single grains \cite{AGHS03}. Hence, in a disordered system
the transport properties will be dominated by the junctions with the highest
gaps. Note also that inelastic cotunneling processes \cite{AN90,GM90}, which
influence the conductance of single junctions at low temperatures, and which
are not considered here, are strongly suppressed in junction arrays.

In conclusion, we find that the gaps observed in conductance measurements on
graphene nanoribbons in Ref. \cite{HOZK07} can be explained as Coulomb gaps
due to the existence of internal junctions between graphene islands, where the
transmission, for all transverse channels, is less than one. We identify the
gaps observed in transport measurements as the effective charging energy of
the islands, renormalized by the charge fluctuations at the junctions. We
further simplify the model by assuming that the transport properties can be
studied by analyzing a single representative junction. The model leads to a
simple dependence of the gaps on material parameters, such as the carrier
concentration, or the width of the ribbons. We obtain a reasonable agreement
with the experiments.

Charging effects are mostly determined by the geometry of the system and by
the amount of screening of the Coulomb interaction. The explanation proposed
here implies that the gaps observed in transport measurements should be weakly
affected by static disorder, or by changes in the electron interference
properties, such as those induced by an applied magnetic field. On the other
hand, we expect that charging effects should be suppressed by metallic gates
at distances from the ribbon which are smaller than the ribbon width.

We are thankful to A. K. Geim, P. Kim, I. Martin, V. M. Pereira, and J. J.
Palacios, for helpful discussions. This work has been supported by MEC (Spain)
under Grants FIS2004-05120, FIS2005-05478-C02-01, the EU Contract 12881
(NEST), the EU Marie Curie RTN Programme no. MRTN-CT-2003-504574, and the
Comunidad de Madrid program CITECNOMIK, no. CM2006-S-0505-ESP-0337. AHCN is
supported through NSF grant DMR-0343790.

\end{document}